\documentclass[conference]{IEEEtran}

\ifCLASSINFOpdf
   \usepackage[pdftex]{graphicx}
   \DeclareGraphicsExtensions{.pdf,.jpeg,.png}
\else
   \usepackage[dvips]{graphicx}
\fi

\usepackage{amsmath}
\usepackage{multirow}
\usepackage{amsmath,amsthm,amsfonts,amssymb,amscd}
\usepackage{lastpage}
\usepackage{enumerate}
\usepackage{fancyhdr}
\usepackage{mathrsfs}
\usepackage{xcolor}
\usepackage{hyperref}
\usepackage{graphicx}
\usepackage{physics}
\usepackage{listings}
\usepackage{hyperref}
\usepackage{array,tabularx}

\usepackage{algorithm}
\usepackage{algpseudocode}
\usepackage{algorithmicx}
\usepackage{slashbox}
\usepackage{textcomp}
\usepackage{amssymb}
\newenvironment{conditions*}
  {\par\vspace{\abovedisplayskip}\noindent
   \tabularx{\columnwidth}{>{$}l<{$} @{${}={}$} >{\raggedright\arraybackslash}X}}
  {\endtabularx\par\vspace{\belowdisplayskip}}

\makeatletter
\newcommand{\newlineauthors}{%
  \end{@IEEEauthorhalign}\hfill\mbox{}\par
  \mbox{}\hfill\begin{@IEEEauthorhalign}
}
\makeatother

\begin{document}

\title{Gait Recovery System for Parkinson's Disease using Machine Learning on Embedded Platforms}

\author{\IEEEauthorblockN{Gokul H.}
\IEEEauthorblockA{\textit{Research Assistant} \\
\textit{Solarillion Foundation}\\
Chennai, India \\
hgokul@ieee.org}
\and
\IEEEauthorblockN{Prithvi Suresh}
\IEEEauthorblockA{\textit{Electronics and Communication} \\
\textit{SRM Institute of Science and Technology}\\
Chennai, India \\
prithvisuresh@ieee.org}
\and
\IEEEauthorblockN{Hari Vignesh B}
\IEEEauthorblockA{\textit{Electronics and Communication } \\
\textit{
SRM Institute of Science and Technology}\\
 Chennai, India\\
havigbaskar@ieee.org}
\newlineauthors
\IEEEauthorblockN{Pravin Kumaar R}
\IEEEauthorblockA{\textit{Computer Science and Engineering} \\
\textit{SRM Institute of Science and Technology}\\
Chennai, India \\
pravin.kumaar99@ieee.org}
\and
\IEEEauthorblockN{Vineeth Vijayaraghavan}
\IEEEauthorblockA{\textit{Director –- Research \& Outreach} \\
\textit{Solarillion Foundation}\\
Chennai, India \\
vineethv@ieee.org}
}

\maketitle

\begin{abstract}

Freezing of Gait (FoG) is a common gait deficit among patients diagnosed with Parkinson's Disease (PD). In order to help these patients recover from FoG episodes, Rhythmic Auditory Stimulation (RAS) is needed. The authors propose a ubiquitous embedded system that detects FOG events with a Machine Learning (ML) subsystem from accelerometer signals . By making inferences on-device, we avoid issues prevalent in cloud-based systems such as latency and network connection dependency. The resource-efficient classifier used, reduces the model size requirements by approximately 400 times compared to the best performing standard ML systems, with a trade-off of a mere 1.3\% in best classification accuracy. The aforementioned trade-off facilitates deployability in a wide range of embedded devices including microcontroller based systems. The research also explores the optimization procedure to deploy the model on an ATMega2560 microcontroller with a minimum system latency of 44.5 ms. The smallest model size of the proposed resource efficient ML model was 1.4 KB with an average recall score of 93.58\%.

\end{abstract}


\section{Introduction}
Freezing of Gait (FoG), defined as a "brief, episodic absence or marked reduction of forward progression of the feet despite the intention to walk" \cite{FoGdefn}, is a gait impairment common among Parkinson's diseased people. According to the survey conducted by Macht et al \cite{macht}, with 6620 people diagnosed with Parkinson's, about 47\% have reported regular freezing, with 28\% of them experiencing FoG on a daily basis.

As FoG is highly sensitive to environmental triggers, cognitive input and medication, it is difficult measure, and as a result, becomes difficult to treat \cite{FoGsens}. Pharmacological treatment of Parkinson's disease with Levodopa (LD) is difficult to manage since it’s effect wears off over time \cite{ld1}. As the disease progresses, more frequent LD administration is required and such gait aberrations often become resistant to pharmacological treatments \cite{ld2}.
\\ Non-Pharmacological treatments include various behavioral ‘tricks’ like marching commands, visual cueing or walking to a music or beat to alleviate FoG symptoms. Lim et al \cite{lim} performed an extensive study on the effects of rhythmic cueing on gait in patients with Parkinson's which suggested an improvement in the walking speed with the help of auditory cues. This treatment consists of $Rhythmic\ Auditory\ Stimulation$ (RAS) \cite{ras} with a metronome producing a rhythmic ticking sound, allowing the patient to recover without external human aid. However, the effects of RAS degrades with time, thereby rendering permanent cueing inadvisable (\cite{fade1}, \cite{fade2}, \cite{fade3}). Thus, auditory cueing is required only in response to occurrences of FoG events. Such a context aware auditory cueing system needs to reliably identify FoG events from normal activities of the patient. With the availability of a wearable, context aware FoG detection system, Parkinson's diseased people can overcome FoG events without external human aid, thus improving their quality of life. 
\\ The authors of this paper propose a scalable sensor based, FoG aware audio cueing system. The system uses a powerful, yet easily deployable Machine Learning algorithm that detects FoG events with high accuracy.
\\In this paper, the architecture of and experiments carried out on the system are organised as follows. Section \ref{syso} elaborates on the different subsystems that are present in the system. Section \ref{meth} explains the implementation of the system and the experiments carried out on a specific subsystem, namely the $ML\ subsystem$. Section \ref{opt} elucidates the optimization of the entire system, for an embedded device with ATMega2560 controller. 

\section{Related Work}
\label{relatedwork}
In order to perform context aware rhythmic auditory stimulation (RAS) during detection of FoG episodes, several investigations have been done using a variety of sensor systems. Delval et al \cite{delval} used data from goniometers worn by Parkinson's diseased people to detect FoG. Nieuwboer et al \cite{neui} analysed electromyographic (EMG)  gait profiles, Handjosono et al \cite{handjosono} studied spatial, spectral and temporal features of electroencephalogram (EEG) signals, while S. Mazilu et al \cite{m15} studied electrocardiography (ECG) and skin conductance from PD patients to design FoG detection systems. However, these sensors are difficult to be integrated as mobile, wearable embedded systems. Thus our research focuses on simple accelerometer-based systems that satisfy the aforementioned requirement. 

Detection of FoG events using accelerometers have been thoroughly worked on in the last decade. Han et al \cite{han}, first attempted to tackle the gait freezing disorder using body worn accelerometers in their monitoring system with frequency domain analysis. Moore et al \cite{moore} made observations on the high frequency components of accelerometers signal (specifically in the 3Hz-8Hz band which was termed as "Freeze band") of the left leg movement during FoG episodes which were not apparent during normal gait ("Locomotor Band": 0.5-3 Hz). They suggested $Freeze\ index$,  which has been widely implemented in this area of research. 
\\Based on the algorithm of Moore et al \cite{moore}, Bachlin et al \cite{bachlin} developed a real time wearable assistant under user dependent settings with a sensitivity of 73.1\% and a specificity of 81.6\% to detect FoG events. However, the algorithm parameters needs to be manually adjusted for optimal results. Jovanov et al \cite{jovanov} performed statistical analysis to set these algorithm parameters which enhanced inter-user adaptability in their real time FoG detection system. Their enhanced implementation on the ARMV7 processor operating at 72 MHz achieved an average latency of 332ms and a maximum latency of 580ms. However, these  FOG detection systems involve a thresholding based algorithm which can be improved by implementing learning algorithms for better performance.\par 
 Machine learning(ML) algorithms are capable of effectively performing accelerometer data based classification tasks, especially in activity recognition (\cite{ar1},\cite{ar2},\cite{ar3},\cite{ar4}). As a result, they were also used in experiments involving FOG detection. Mazilu et al \cite{mazilu} proposed a FoG detection system capable of being deployed in smart phones with average specificity and sensitivity of more than 95$\%$ followed by Oung et al \cite{oung}. Though ML methods have exceptional performance in this domain, the model’s size and computational complexity requirements limits the scalability of these systems. They can be deployed as cloud based IoT systems but high operational latencies due to data transmission are inevitable.

These systems can be made much more pervasive, mobile and cheaper for the consumers with the implementation of lightweight learning algorithms that can be deployed in resource constrained, embedded devices without compromising the performance. In this context, a $resource\ constrained\ device$ or an $edge\ device$ refers to an embedded device with less than 32KB RAM and 16Mhz processor. The authors of this paper, implement and study two such lightweight learning algorithms called  $ProtoNN$ by Gupta et al \cite{protonn} and $Bonsai$ by Ashish et al \cite{bonsai}.
 
\emph{ProtoNN} is a novel, k-nearest neighbors (kNN) based general supervised learning algorithm that can be deployed on tiny edge devices, whilst maintaining state-of-the-art accuracies for typical IoT prediction tasks, with just about 16KB of memory size. 
\emph{Bonsai} is a new tree model for supervised learning tasks such as binary and multi-class classification and can be deployed on small embedded devices. Bonsai can also fit in the L1 cache of processors found in mobiles, tablets, laptops and servers for low-latency applications.

\section{Problem Statement}
\label{ps}
Although existing Machine Learning (ML) models provide good solutions for detection of FoG, implementation and scalability of these models on embedded platforms are difficult owing to their limitations, as stated in Section \ref{relatedwork}. During real time application, an FoG event lasts only for a few seconds. Hence the system must be capable of operating with less computational time, as a lower latency implies capability of faster and seamless recovery via RAS.

Thus, the authors of this paper propose an assistive system consisting of a Machine Learning subsystem capable of detecting such events on-device. This avoids the data transmission latency that is otherwise present in cloud based ML-IoT systems. The proposed system provides high accuracy, comparable to traditional Machine Learning Classifiers, on embedded devices with extremely small model size and inference time. This is achieved by extracting powerful features from signals of the accelerometers, with computational time in the order of a few milliseconds. These features are passed to a lightweight Classifier that detects the FoG event. The system was optimised to be  compatible on an ATMega2560, reifying the scalability of the system across wider range of embedded platforms.

\section{System Architecture}
\label{syso}
\begin{figure}[!ht]
\centering
\includegraphics[width = \columnwidth]{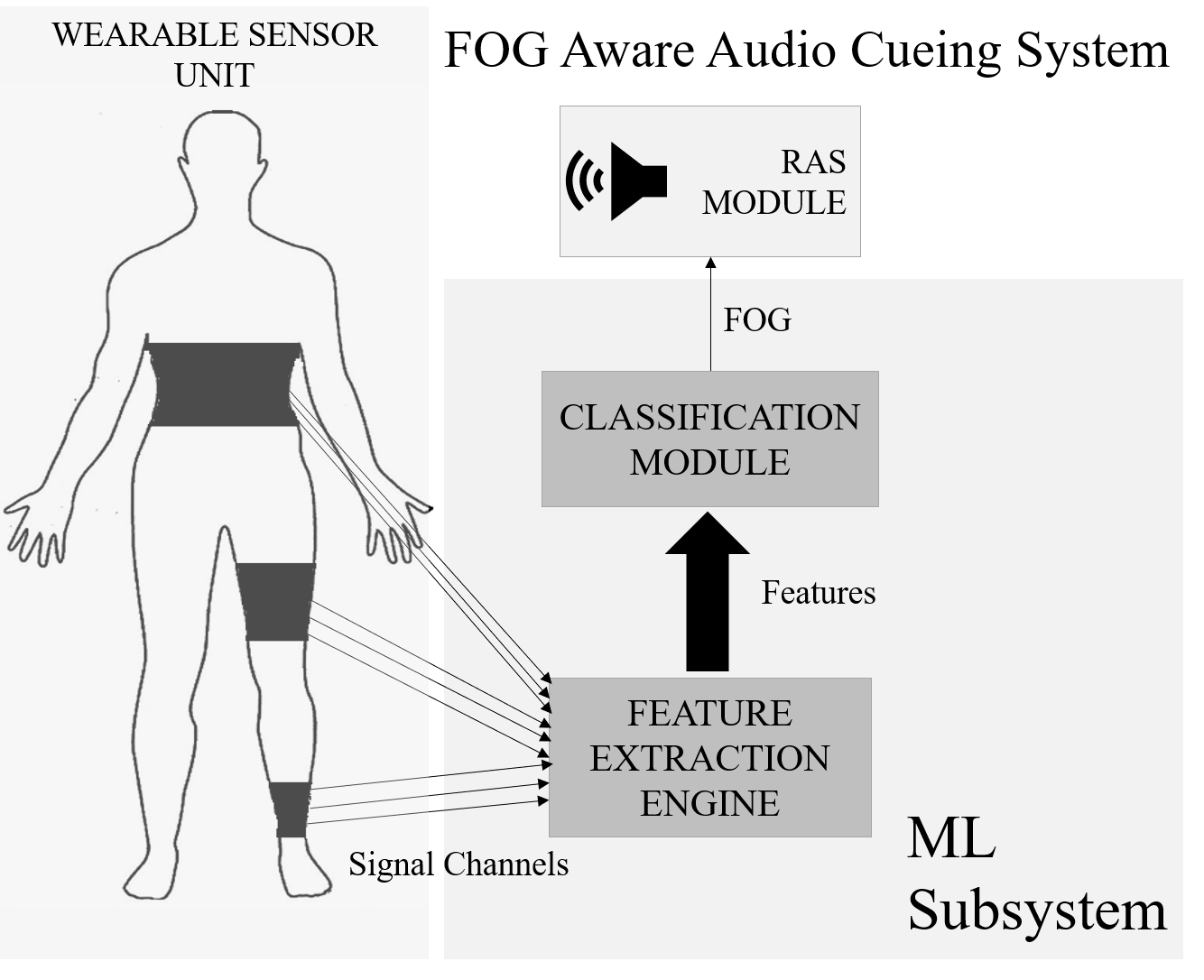}
\caption{System Architecture}
\label{overview}
\end{figure}
The proposed FoG aware auditory-cueing system obtains signals from the
\textbf{wearable sensor unit} and is fed into the pre-processing stage of the \textbf{ML subsystem} termed as the $Feature\ Extraction\ Engine$.
The $Feature\ Extraction\ Engine$ extracts necessary signal features and is pipelined to a (ML) $classifier$ which detects the activity. Upon detection of an FoG event, an output signal is sent to the \textbf{RAS module} to initiate recovery from FoG. These system components are further elaborated as follows.

\subsection{Wearable Sensor Unit}
This input unit consists of 3 tri-axial accelerometers placed on the Ankle (A), Leg(L) and Torso(T). Each channel constitutes of a signal corresponding to a \emph{single axis} from an \emph{accelerometer}. Thus a total of 9 channels are present and they are represented by the set $\Gamma $ as shown,
\begin{equation}
    \Gamma = \{A_X,A_Y,A_Z,L_X,L_Y,L_Z,T_X,T_Y,T_Z\}
\end{equation}

\subsection{ML subsystem}
\label{mlsub}
\subsubsection{Feature extraction engine}
\label{fee}
From the signals of the sensors mounted, the $Feature\ Extraction\ Engine$ extracts features from each channel ($\gamma, where\ \gamma \in \Gamma$) and passes on these features to a machine learning $classifier\ module$ to detect the activity. In this pre-processing stage, the data received from each channel is stored as a 1-dimensional window of $w\times f_s$ time-steps, where $w$ is the length of the window in seconds and $f_s$ is the sampling frequency in Hz. 

\begin{equation}
    \gamma = \left[ \begin{array}{cccc}
         \gamma_1 & \gamma_2 & ... & \gamma_{w\times f_s}  \\
    \end{array}\right]_{w\times f_s}
\end{equation}

In this research, 5 time domain and 5 frequency domain features, whose description is elucidated in Table I, are extracted from each window for all the channels forming a feature set ($F_D$).

The time domain features include mean ($\mu$), standard deviation ($\sigma$), variance ($\sigma^{2}$), root mean square ($rms$) and mean absolute value ($mav$). Among the frequency domain features, Freeze Index ($FI$) which was first proposed by Moore et al \cite{moore}, is found to be an important feature in existing works (\cite{m15},\cite{moore},\cite{bachlin},\cite{mazilu},\cite{oung}) mentioned in section \ref{relatedwork}.  Other frequency domain features include Band Power($P$), Energy($E$), Entropy($S$) and Peak Frequency($f_{peak}$). To eliminate redundancies and avoid over-fitting, the most discriminating features from the available set of features are selected to form a smaller subset ($F_d$) which will be used by the classification module.
\begin{table*}[]
\label{featuretable}
\centering
\caption{Feature set ($F_D$) }
\begin{tabular}{|l|l|l|l|}
\hline
\textbf{Time Domain Features}             & \textbf{Description}  \\ \hline
Mean                &     The average value of the signal in the window   \\ \hline
Standard Deviation            &  Deviation of the signal in a window as compared to its mean value.           \\ \hline
Variance  &  Square root of Standard Deviation.           \\ 
\hline & \\[-0.5em]
Root Mean Square    &  Square root of mean of the squared signal in the window:  \\  & $\sqrt(\frac{1}{N}\sum_{n=1}^N x_n^2)$           \\ [0.5em]
\hline & \\[-0.5em]
Mean absolute Value &  Mean of absolute of the signal in the window: \\
&$\frac{1}{N}(\sum_{n=1}^N \mid x_n \mid )$          \\[0.5em]
\hline & \\[-0.5em]
\textbf{Frequency  Domain Features}             & \textbf{Description}  \\ \hline
Entropy                &     The Measure of the random distribution of frequency components   \\ \hline
Energy            &  The sum of squared magnitudes of FFT components of the signal, divided by window length.           \\ \hline
Peak Frequency  &  Value of maximum frequency in the power spectrum.           \\ 
\hline & \\[-1em]
Freeze Index    &  Power of signal in freeze band (3-8Hz) divided by it's Power in locomotor band(0.5-3Hz)  : \\
& $\frac{\sum_{j=3}^{j=8}P_j}{\sum_{j=0.5}^{j=3} P_j}$, where P denotes Power   \\ [1em]
\hline & \\[-0.5em]
Band Power &  Sum of the power in freeze band and in locomotor band :    ${\sum_{j=3}^{j=8} P_j} + {\sum_{j=0.5}^{j=3}P_j} $     \\ [0.5em]
\hline

\end{tabular}
\end{table*}

\subsubsection{Classification Module}
\label{classification}
The Classification module of the ML subsystem consists of a Machine Learning algorithm, that detects whether a person is undergoing FoG or not. The module takes the features from the Feature Extraction Engine as the input, passes it through the algorithm and sends a signal to the RAS module, if the person is undergoing FoG.\\
The detection problem is modelled as a binary classification problem with an occurrence of FoG as the positive class. In this binary classification problem, the recall score of the positive class is also termed as $sensitivity$ while the recall score of negative class is termed as $specificity$. These metrics which are extensively found in the literature and used throughout this paper are given by the formula,
\begin{equation}
\label{metrics1}
Sensitivity = \frac{TP}{TP+FN}
\end{equation}   
\begin{equation}
\label{metrics2}
Specificity = \frac{TN}{TN+FP}
\end{equation}   
where,
\begin{conditions*}
TP & True\ positives\ $FoG$ events classified as $FoG$
events\\
TN & True\ negatives\ $Normal$ events classified as $Normal$ events\\
FP & False\ positives\ $Normal$ events classified as $FoG$ events\\
FN & False\ negatives\ $FoG$ events classified as $Normal$ events
\end{conditions*}
For this system, various Machine Learning classifiers were experimented with, keeping in mind factors such as accuracy, latency and deployability including the best performing benchmark models proposed by Mazilu et al \cite{mazilu} (Random Forests) and Oung et al \cite{oung} (Support Vector Machine).  The results and inferences of these Machine Learning classifiers are presented in section V. 

\subsection{RAS module}
The RAS module placed in the patient's ear consists of a circuitry capable of emitting auditory cues and is triggered based on the signal from the classifier model. The RAS module, when triggered, sends out metronome click-embedded music that helps people undergoing FoG to regain control of their gait independently.

The run time implementation and experimentation of this system is performed in section \ref{meth}.

\section{System Implementation}
\label{meth}

\subsection{Dataset}
The run-time experiments were carried out on the DAPHNet dataset \cite{bachlin}, which is publicly available. This dataset was a collaborative effort of  the Laboratory for Gait and Neurodynamics, Tel Aviv Sourasky Medical Center, Israel and the Wearable Computing Laboratory, ETH Zurich, Switzerland. Recordings were run at the Tel Aviv Sourasky Medical Center in 2008. The study was approved by the local Human Subjects Review Committee, and was performed in accordance with the ethical standards of the Declaration of Helsinki. The dataset consists of accelerometer data obtained from 10 patients who are diagnosed with Parkinson's disease. 3 Body worn accelerometers, on the ankle, leg and torso were attached to the the patient while they were asked to perform the following activities :
\begin{itemize}
    \item Walking back and forth with 180\textdegree\  turns
    \item Walking with a series of initiated stops and 360\textdegree\  turns 
    \item Walking resembling activities in daily life (ADL), such as entering rooms, making coffee or leaving rooms, etc.
\end{itemize}
The data was sampled at a sampling frequency ($f_s$) of 64Hz. In total, 8 hours and 20 minutes worth of data was collected, in which 237 events of FoG have been labelled. The duration of the FoG events were in between 0.5s to 40.5s, with a mean of 7.3s and standard deviation of 6.7s. Over 50\% of the FoG events lasted longer than 5s and 93.2\% of the FoG events were less than 20s. Synchronized video recordings were taken and analyzed by physiotherapists to label these FoG events. There were three classes that were labelled namely, 0-debriefing, 1-Normal activity and 2-Freezing Of Gait. 
The data was provided for ten users as separate set of files. Each line of the file consisted of 1 sample of data with the following fields:

\begin{enumerate}
    \item Time Stamp
    \item Accelerometer in the ankle's X axis 
    \item Accelerometer in the ankle's Y axis.
    \item Accelerometer in the ankle's Z axis.
    \item Accelerometer in the Leg's X axis.
    \item Accelerometer in the Leg's Y axis.
    \item Accelerometer in the Leg's Z axis.
    \item Accelerometer in the Torso's X axis.
    \item Accelerometer in the Torso's Y axis.
    \item Accelerometer in the Torso's Z axis.
    \item Label indicating FoG or Walking or ADL
\end{enumerate}

Out of the 11 $fields$ provided, only the 9 columns/channels pertaining to the accelerometer readings were used in the Feature Extraction Engine. Out of the 10 patients, patient 4 and patient 10 did not undergo FoG during the experiment procedure. Hence, the data from these two patients were excluded from our experiments.

\subsection{ML subsystem - Implementation}
\label{Implementation}
\begin{figure*}[!ht]
\centering
\includegraphics[width = 7.45in
]{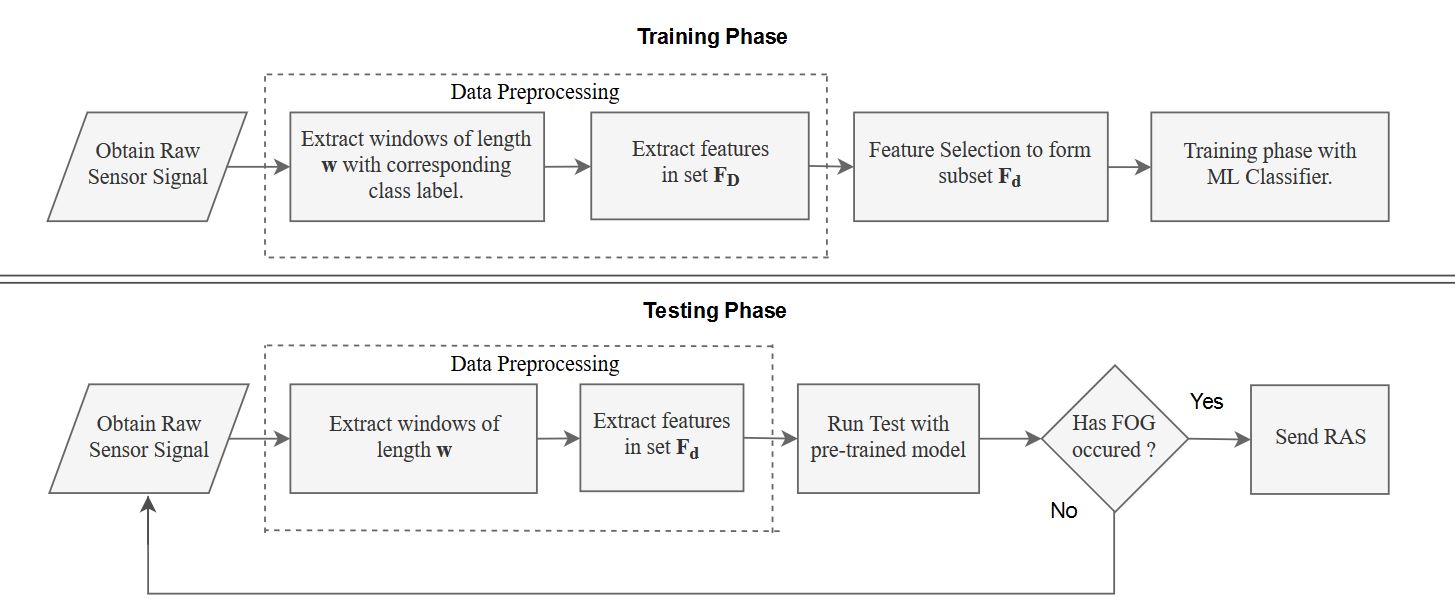}
\caption{Flowchart of Training and Testing phase}
\label{ttf}
\end{figure*}

From the dataset, the 0-debriefing class was omitted because they were not part of the experiment, allowing for binary classification between $FoG$ (positive class) and $Normal$ (negative class). Various Machine Learning classifiers were implemented in the classification module for reasons stated in subsection \ref{classification}. The train-test split ratio of the dataset that is fed to the models is 70\%-30\%. The general training and testing process of this ML based FoG detection system implementation is represented in Fig. \ref{ttf}.

The classifiers that were implemented are categorized as 
\begin{enumerate}
    \item \textbf{Standard ML classifiers}:
    
    \begin{table*}[]
\caption{Recall scores for Subject dependant test- Standard ML models}
\centering
\label{stdmlresults} 
\begin{tabular}{|c|c|c|c|c|c|c|c|c|c|}
\hline
\setlength{\arrayrulewidth}{0.1mm}
\setlength{\tabcolsep}{10pt}
\renewcommand{\arraystretch}{1.1}
\multirow{2}{*}{Classifier} & \multirow{2}{*}{Size (KiloBytes)} & \multicolumn{2}{c|}{w=1} & \multicolumn{2}{c|}{w=2} & \multicolumn{2}{c|}{w=3} & \multicolumn{2}{c|}{w=4} \\ \cline{3-10} 
 &  & Spec($\%$) & Sens($\%$) & Spec($\%$) & Sens($\%$) & Spec($\%$) & Sens($\%$) & Spec($\%$) & Sens($\%$) \\ \hline
\textbf{DecisionTree} & \textbf{109.57}& 95.98& 91.28& 95.73& 92.95& 98.16& 96.07& 98.92& 96.70  \\ \hline
\textbf{Random Forest} &\textbf{545.04}&	97.92	&98.51&	98.95&	98.56&	99.76&	98.79&	99.44&	98.91 \\ \hline
AdaBoost & 946.44& 97.55& 96.03& 98.62&96.57&	99.29&	97.86&	99.56&	97.99  \\ \hline
K Nearest Neighbors &13005.55&	99.67&	86.087&	98.83&	90.35&	96.97&	93.56&	97.38&	94.61  \\ \hline
Support Vector Machine &2966.57&	94.65&	90.96&	96.15&	93.98&	97.54&	95.86&	98.64&	97.547  \\ \hline
\end{tabular}
\end{table*}

    These classifiers consist of ML models with the following characteristics,\begin{itemize}
        \item Large space in memory: These models occupy over 100 KB, thus taking up a sizable portion of memory.
        \item Computationally intensive: These models require large RAM,  for inference. This is due to reasons such as the High Dimensional Feature space, retaining entire training set in memory during training, etc
    \end{itemize}
    
    The classifiers specified in Table \ref{stdmlresults} were tested out in the classification module and the results of 10 fold cross validation for these experiments are portrayed in the same. These classifiers were trained on the data from features extracted from the signals as described in section \ref{fee} for varying window lengths, $w$ $\in$ $\{1,2,3,4\}$. Since the maximum efficiencies were observed at $w$ = 4 and the system latency  increases with higher values of $w$, we restrict our maximum value of $w$ in our experiments to 4 (256 timesteps).  
    
\begin{table*}[]
\label{edgeresults}
\centering
\caption{Recall scores for subject dependant tests- Resource efficient ML Models}
\begin{tabular}{|c|c|c|c|c|c|c|c|c|c|}
\hline
\setlength{\arrayrulewidth}{0.1mm}
\setlength{\tabcolsep}{10pt}
\renewcommand{\arraystretch}{1.1}
\multirow{2}{*}{Classifier} & \multirow{2}{*}{Size (KiloBytes)} & \multicolumn{2}{c|}{w=1} & \multicolumn{2}{c|}{w=2} & \multicolumn{2}{c|}{w=3} & \multicolumn{2}{c|}{w=4} \\ \cline{3-10} 
 &  & Spec($\%$) & Sens($\%$) & Spec($\%$) & Sens($\%$) & Spec($\%$) & Sens($\%$) & Spec($\%$) & Sens($\%$) \\ \hline
ProtoNN    & 2.92           & 94.83      & 90.44     & 97.65      & 91.82     & 98.82      & 94.57     & 99.66      & 95.25      \\ \hline
Bonsai     & 6.71           & 94.04      & 88.27     & 95.79      & 90.05     & 96.71       & 91.76      & 98.36      & 92.9       \\ \hline
\end{tabular}
\end{table*}
    \item \textbf{Resource-efficient ML classifiers}: 
	These classifiers, in contrast to the standard ML classifiers, were crafted to occupy low space in memory and are computationally light. These characteristics allow these classifiers to be deployed on devices that have RAM in the order of a few kilobytes. These Resource efficient ML classifiers also  were observed to have a small test inference time of the order of 20ms per data point. These models project the training data into a lower dimensional space where prediction is made simpler, reducing model complexity and size, resulting in faster computation. 
	
The resource efficient classifiers used by the authors in this paper are Bonsai \cite{bonsai} and ProtoNN \cite{protonn}. The results of the classifiers under similar experimental conditions as that of the Standard ML classifiers are portrayed in Table III.

\end{enumerate} 
As observed from Table \ref{stdmlresults}, all standard ML classifiers perform almost equally well for a given window size. Whilst considering deployability of the model on a resource constrained device, \textbf{Random Forests and Decision Tree} have comparable metric scores with relatively smaller model size. Provided there is scope for further compression of these models without reducing the training data, these two classifiers were chosen for optimization and were pit against the two resource efficient classifiers, \textbf{ProtoNN and Bonsai}. Section \ref{opt} elaborates on selecting the most suitable out of these four classifiers for a deployable ML model in a resource constrained device. 

\section{System Optimization and Results}
\label{opt}
The optimization of this gait recovery system is targeted to achieve our suggested minimum-level configuration: to be deployed on a 16MHz Arduino Mega which uses an ATMega2560 micro-controller with an 8KB internal SRAM. For the ML subsystem to even theoretically fit such low specifications, the ML classifier model's size needs to be constrained within few kilobytes. This optimization is done by studying the size vs performance characteristics of the four models suggested in section \ref{Implementation} which is elaborated in \ref{mopt}.

\subsection{Model Optimization}
\label{mopt}
Decision Tree and Random Forests are generally categorized as Tree-based classifiers. The size of these tree based classifiers can be reduced by the process of tuning it's hyper-parameters such as Depth. This process to reduce model size and prevent over-fitting is termed as $pruning$ \cite{prune}. 

The hyper parameters of ProtoNN; projection dimensions ($\hat{d}$) and number of prototypes (m) were tuned, in accordance to the binary class implementation of ProtoNN by Gupta et al. \cite{protonn}, to achieve lower model sizes. Similarly, the parameters of Bonsai; projection dimension and depth were altered to minimise it's model size \cite{bonsai}. The performance of these models with varying model sizes is shown in Fig 3.

Though higher recall scores were achieved by Decision Tree and Random Forests at optimal settings as seen in Table \ref{stdmlresults}, these models performed poorly in our tests under compressed settings as inferred in Fig 3. At a model size of 1.4KB, the average recall score of Decision Trees was at 66.4\%  while that of ProtoNN and Bonsai was at 92.3\% and 91.6\% respectively, with ProtoNN marginally outperforming Bonsai.  The least achievable model size of the Random Forests ensemble model, was at 7.2KB  with an average recall score of 68.9\%. This is attributed to the explicit imposition of sparsity constraints on the parameters by resource-efficient models during the training optimization itself to obtain an optimal model within the given model size de-facto, instead of post-facto pruning by the standard models to force the model to fit in memory \cite{protonn}. Owing to it's practical viability in real-time \cite{gesturepod} and better performance under the minimum setting configuration, we choose ProtoNN as our model to deploy on the ATMega2560.  

\begin{figure}[!ht]
\centering
\label{pvsd}
\includegraphics[width = 3.3in
]{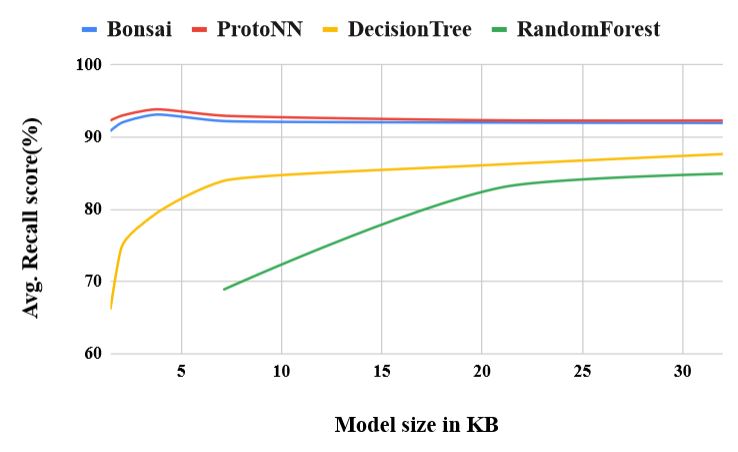}
\caption{Model size vs Recall score for ProtoNN and DecisionTree}
\label{circuitdiagram}
\end{figure}

\subsection{Feature set Optimization}
Though ProtoNN model can fit in less computationally powerful embedded devices, in order to deploy the ML subsystem in an ATMega2560, there is a need to alter the extracted features to satisfy low computational complexity and latency of the Feature Extraction Engine. However, these features must also be capable of ensuring good classification performance. This is possible in this FoG detection problem since the feature set $F_d$, after the feature selection process mentioned in Section \ref{fee}, involved a considerable number of easily computable features, mostly from the time domain.

The time domain subset of the feature set $F_D$(=90 features) is denoted as $F_{TD}$ (=45 features). From $F_{TD}$, the redundant features are eliminated to form a small subset $F_{td}$. With ProtoNN as the classifier of this ML subsystem, the average recall scores with the "all-feature" subset $F_d$ and the currently optimised time domain feature subset $F_{td}$ are compared in Table \ref{fdvsftd_ms} with their corresponding computational time (in ms) when tested in Arduino Mega (ATMega2560), for all values of $w$.

\begin{table}[h!]
\centering
\caption{Average computational time(ms) and Recall score comparison between feature subsets $F_d$ vs $F_{td}$}
\label{fdvsftd_ms}
\setlength{\arrayrulewidth}{0.1mm}
\setlength{\tabcolsep}{10pt}
\renewcommand{\arraystretch}{1}
\begin{tabular}{|c|c|c|c|c|}
\hline
\multirow{2}{*}{\textbf{\begin{tabular}[c]{@{}c@{}}$w$ (sec)\end{tabular}}} & \multicolumn{2}{c|}{\textbf{\begin{tabular}[c]{@{}c@{}}Computational\\ Time (ms)\end{tabular}}} & \multicolumn{2}{c|}{\textbf{\begin{tabular}[c]{@{}c@{}}Average Recall\\ Score (\%)\end{tabular}}} \\ \cline{2-5}
 & \textbf{$F_d$} & \textbf{$F_{td}$} & \textbf{$F_d$} & \textbf{$F_{td}$} \\ \hline
\textbf{1} & 354.1 & 24.5 &92.64  & 91.23  \\ \hline
\textbf{2} & 747.5 & 49.3 & 94.73 &93.58  \\ \hline
\textbf{3} & 1105.3 & 79.1 & 96.70 & 94.63 \\ \hline
\textbf{4} & 1471.2 & 100.5 & 97.45  &95.77 \\ \hline
\end{tabular}
\end{table}

While including frequency domain features in our model yields better recall score, it can be inferred from the Table \ref{fdvsftd_ms} that the latency due to these features compared to it's time domain counterpart is very high. This is attributed to the simplicity of time domain features which are based on simple mathematical and statistical operations whereas the frequency domain needs performance of Fast Fourier Transform (FFT). Computing FFT of even a single signal channel window increases the computational complexity and also demands the use of complex variables for storing the real and imaginary part of the transformed signal, thereby increasing the subsystem's memory requirements. 

The total inference time of the ML subsystem is the sum of latencies of the Feature Extraction Engine and the classification module. The ProtoNN classification module predicts the class within 20ms, on an average. The minimum Feature Extraction Engine latency is for $w$ = 1, as observed in Table IV. This makes the least total inference time of the ML subsystem at 44.5 ms. 

However, this small inference time is at the cost of compromising the accuracy. Thereby, the optimum window length, $w_{opt}$, proposed is 2 seconds (128 time steps), with an average recall score of 93.58\%. 

This small size, low latency ML subsystem can now be effectively deployed in the ATMega2560 microcontroller which is considered as our minimum level configuration. The system's complexity can be further reduced by optimising the number of sensors used, also enhancing the wearablity of the device.

\subsection{Sensor Optimization}
\begin{figure}[h!]
\centering
\label{sensor}
\includegraphics[width = 3.45in, height = 2.5in
]{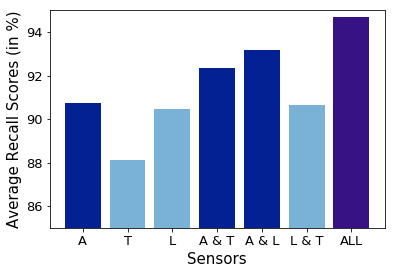}
\hspace{-3mm}
\caption{Accuracy scores for various Sensors}
\end{figure}

We recall that the three accelerometers used in this system are positioned on different parts of a person's body namely, Ankle(A), Leg(L) and Torso(T). Every accelerometer has 3 axes of signal channels, with a total of 9 channels. Using all the 3 sensors for classifying increases the system complexity due to larger number of data dimensions resulting in a larger feature set. Hence, The system's average recall scores for  various permutations of the sensors at $w_{opt}$ = 2 were analysed to infer the importance of various sensors in the \textit{Wearable Sensor Unit}, depicted in Figure \ref{sensor}.

Though the system performs best with all the three sensors, it can be inferred from Figure \ref{sensor} that eliminating the torso sensor does not affect the system's performance considerably. This effect was confirmed by observing feature set $F_{td}$, wherein only 2 features corresponding to the torso were present. In addition to this, the model trained using the ankle sensor alone yields higher recall scores than the model trained on the torso and leg individually. By making optimal trade-offs in usage of sensors, the system complexity and latency is further reduced with a smaller yet powerful feature set. Thus, we eliminate hardware redundancies while also improving the system's performance. 

\section{Conclusion}
In this paper, system optimization of a Freezing of Gait (FoG) aware audio cueing system was carried out with the implementation of a resource efficient Machine Learning model. Our proposed system involving this low-sized model, ProtoNN, requires 1.4 KB of space with a specificity and sensitivity of 97.65\% and 91.82\% respectively, for $w_{opt}$, notwithstanding the meagre 1.3\% drop in average recall score as compared to standard ML classifiers. A suitable choice of less computationally demanding time domain feature set facilitated deployment of the Machine Learning subsystem in a basic embedded micro-controller based platform, the Arduino Mega with ATMega2560 consisting of just an 8KB RAM. In the ATMega2560, the average time taken by this system to extract feature set $F_d$ is 747ms and to detect the class is 20ms for $w_{opt}$. This feature extraction time, for $w_{opt}$, was reduced to 49.3ms for the optimized feature set $F_{td}$, thereby reducing the system latency and complexity with a minimal decrease in it’s efficiency. The average recall score for this system is 93.58\%. By trading the number of sensors used, the system complexity was further reduced to fit in the ATMega2560. By deploying this system in a micro-controller at a minimum level configuration, we conclude that the system can be scaled for a wider range of embedded devices.

\section{Future Work}
In the future, this system can be tweaked to perform FoG prediction, which helps in detecting the FoG event much before its onset. This FoG prediction problem can be solved efficiently with the use of recurrent neural networks(RNN) \cite{torvi} and ML models  with a powerful feature set \cite{orphan} which involves much higher system complexity. However, there is scope to resolve the model complexity of RNNs by implementing methods like early prediction and multiple instance learning \cite{emi} in order to be deployed on embedded devices.

\section*{Acknowledgment}
The authors would like to acknowledge Solarillion Foundation for its support and funding for this research work.

\end{document}